# Optical image-based thickness characterization of atomically thin nanomaterials using computer vision techniques

Daniel Cui

## Abstract

The main objective of this study was to develop a novel method of characterizing nanomaterials based on the number of layers without the aid of state-of-the-art electron and force microscopes. While previous research groups have attempted to establish a correlation between optical image contrast and layer number for inferring layer numbers of nanomaterials with already well-known software such as ImageJ and Gwyddion SPM Analysis, the work for this study strived to automate the image contrast-based characterization of the layer numbers using computer vision algorithms. After acquiring the necessary data points using graphene samples from another study and nanoscale $MoTe_2$ samples through an experimental method consisted of using both ImageJ and Gwyddion, curve fitting in RStudio was used to create quadratic models that were incorporated in a computer vision method composed of three algorithms. In total, 12 $MoTe_2$ samples and 16 graphene samples were used in 28 test trials in order to determine the algorithms' efficiencies in layer number characterization. Ultimately, a success rate of 89% was obtained with an average overall run time of 15 seconds for the computer vision method. As a consequence, this computational method may be faster, more effective, and more cost-effective than current widely used electron, atomic force, and optical microscopy techniques.

## **Introduction**

With the influx of numerous advances in microscopy, the visualization and characterization of materials beyond the boundaries of the human eye have opened up new and untouched realms on the microscopic and even the atomic scale. Most notably, the advent of electron microscopes such as the scanning electron microscope (SEM) and the transmission electron microscope (TEM) have allowed scientists to investigate nanoscale media (Fig. 1b) utilizing the quantum tunneling of electrons on the nanoscale, dominated by quantum mechanical forces that are just only being elucidated now (*1*). Additionally, force microscopes such as the atomic force microscope (AFM) have assumed instrumental roles in probing the nano world through the use of vibrating metal tips (Fig. 1a). Additionally, a multiple beam interferometry method developed by Scott et al. demonstrated substantial capability in characterizing nanomaterial thickness as well (*2*). Without these revolutionary instruments, research in nanoscience and nanotechnology, which has already spawned a multitude of innovations such as semiconductors (Fig. 1c), catalysis for green energy applications, and novel cancer treatments (*3*), would not be possible and as groundbreaking as it is today. The instruments can characterize molybdenum ditelluride and graphene, the two materials used in this study. They have immense implications in nanoelectronics and nanophotonics, the study of the behavior of light on the nanoscale, for creating semiconducting transistors and radiation detectors (*4, 5*).

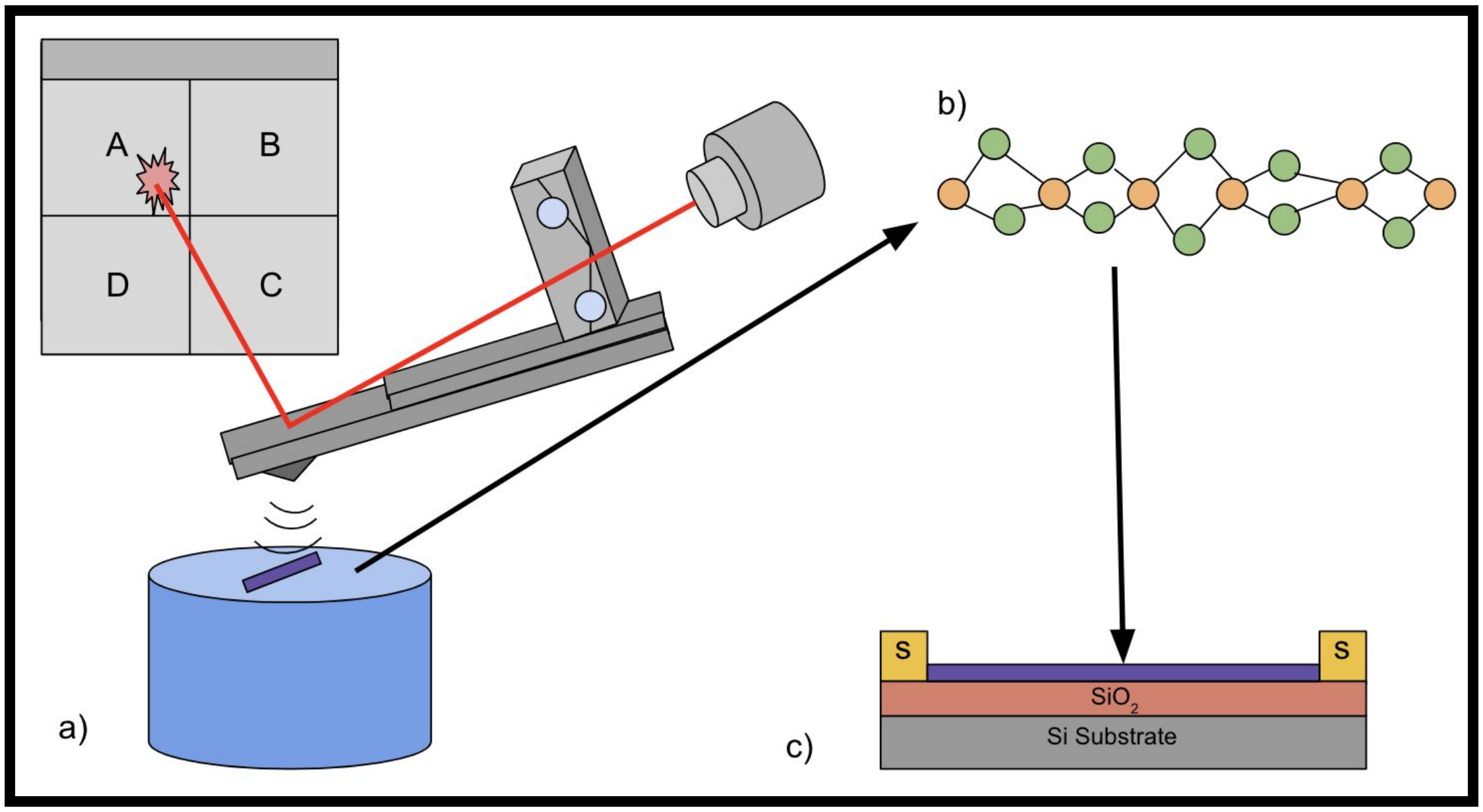

**Figure 1a, 1b, 1c**: **1a)** Illustration of a vibrating AFM tip powered by piezoelectric motors scanning over nanoscale media. A laser shining over the tip and a detector are used to translate the deflections in the tip into an image. **1b)** the atomic structure of a hypothetical nanomaterial. **1c)** a potential application of nanomaterials in a gated semiconductor with a source and sink on the left and right sides respectively. The silicon dioxide ($SiO_2$) and silicon substrates are mainly used for insulation.

Nevertheless, microscopy tools such as the SEM, TEM, AFM, and multiple beam interferometry have substantial flaws in terms of their functionality. Despite the fact that they are capable of outputting high-resolution images of materials that may only be a few atoms thick, electron and force microscopes 1) consume a significant amount of time and monetary resources and 2) suffer considerable performance issues from disruptive interferences and vibrations (*6*). Since most nanoscale imaging tools require inert conditions within a vacuum chamber, stable voltages and currents, and a robust water-cooling system, operating the apparatuses can be tedious and also requires a substantial amount of training in order to acquire the necessary knowledge and skills (*7*). Furthermore, because electron and force microscopes are not only sophisticated pieces of equipment but also necessitate a stable environment for effective and accurate measurements, the cost of acquiring and using such high-quality tools can easily surpass one million dollars (*7*). For instance, vibrations from roadwork near the laboratory or external magnetic fields can disrupt the voltage and current that are needed for effective performances. Furthermore, while optical based methods of characterization, such as multiple beam interferometry, improve in their abilities to resist environmental interferences and operation times (*3*), the setups that are required can be extremely costly as well and usually have to be executed in atmosphere, rendering analyses of degradable materials ineffective.

In the past few years, researchers have been endeavoring to find more efficient and cost-effective methods of nanomaterial characterization. One of these attempts has been mainly rooted in correlating optical image contrast measured using a confocal microscope with the thickness of a sample using electron or force microscopes in order to determine a correlation between sample image contrast and thickness (*8-11*). This correlation could then be used to infer thicknesses of other nanomaterials solely based on their optical images without the use of electron or force microscopy. Therefore, this study endeavors to expand and improve upon the effort in using image contrast to characterize nanoscale matter by developing a fast and automated process based on computer vision that accomplishes the characterization task. As a consequence, the main goal was to be able to input optical images and output the layer number and visualizations of the sample's layers with considerable computational efficiency without the aid of high-powered tools such as SEMs, TEMs, AFMs, or multiple beam interferometry. Though some disadvantages of current characterization methods, such as the enforcement of inert conditions in a vacuum chamber and susceptibility to environmental interferences, were unavoidable in the experimental methods, the aim ultimately was to use the data gathered using standard experimental methods to develop a novel computer vision method. Therefore, the significance of this work lies in the ability to predict the thickness of nanomaterials using a computational characterization method rather than the physical process of generating thickness information.

## Materials and Methods

### *Experimental Materials and Methods*

In order to effectively develop a set of computer vision algorithms to characterize nanomaterial thicknesseses, a standard experimental method was first used in lab to acquire contrast and thickness data sets of 8 nanoscale crystal samples. Specifically, the optical microscope method was used to acquire the contrast values, while the Gwyddion method measured the layer number of the nanomaterials (Fig. 2).

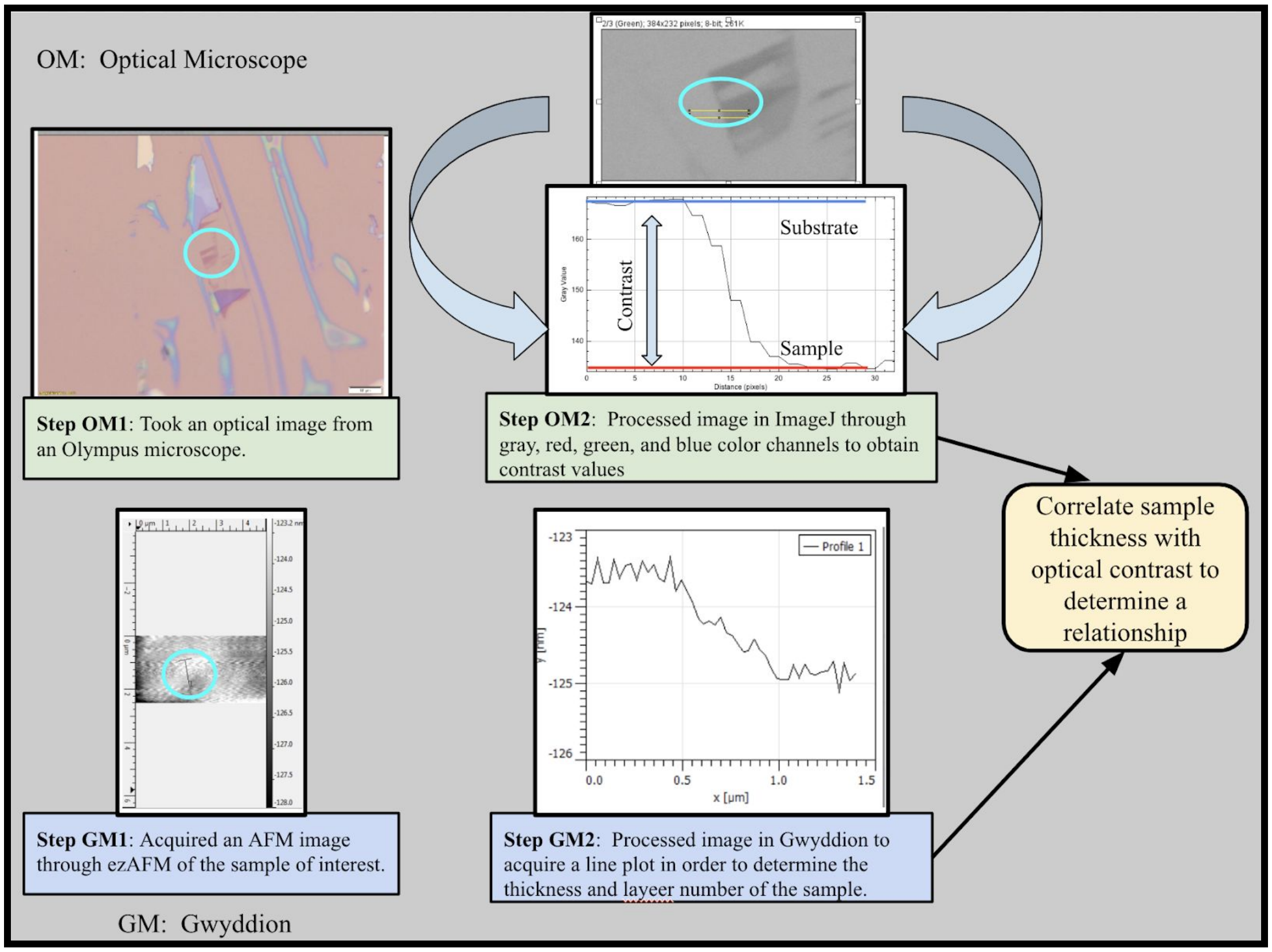

**Figure 2**: Diagram showcasing the steps taken to obtain optical contrast values for the gray, red, green, and blue color channels using an optical microscope and ImageJ (*12*) and layer numbers using an AFM and the Gwyddion SPM analysis software (*14*).

Using the Olympus BH2 Fluorescence Microscope in a vacuum chamber, digitally colored pictures of 8 different samples of nanoscale molybdenum ditelluride ($MoTe_2$) were taken on a 300 nm silicon substrate at a magnification of 100X and exposure time of 50 ms. Once these pictures were stored as PNG files on the computer, a public domain image-processing program, developed by the National Institutes of Health called ImageJ (*12*), was then used to zoom in on 12 different uniform regions of interest on the $MoTe_2$ samples, convert the zoomed-in image into 8-bit grayscale or RGB stack images, and obtain the line profile of the pixel intensity values based on the gray, red, green, and blue color channels using ImageJ's plot profile function. The list of x and y data points from the line profile graph was then extracted and inputted into a Google Sheets file. The x ranges for the substrate and sample pixel values were specified based on the line profile so that an averaging function could be applied. The averages of sample pixel values for each grayscale color channel were then subtracted from the averages of those of the substrate to obtain the gray, red, green, and blue contrast values.

To determine the thicknesses of the $MoTe_2$ samples, an atomic force microscope made by Nanomagnetics Instruments paired with the ezAFM software (*13*) was utilized to acquire AFM images of the 8 samples. These images were then transferred to the Gwyddion SPM Analysis software platform (*14*) in order to remove noise and zoom in on specific areas of interest. Using the plot profile function in the software, a line was drawn across the boundary between the sample region and the silicon substrate to output a line profile graph with nanometers on the y-axis and micrometers on the x-axis. Ultimately, the difference was taken between the y-coordinates for the two relatively smooth and straight regions of the line profile in order to obtain layer thicknesses of each of the 12 regions. These layer thicknesses were then converted into layer numbers based on the fact that each layer of $MoTe_2$ was about 0.7 nm thick (*15*).

*Computational Materials and Methods*

Based on the contrast and layer number data acquired through the experimental method, three separate computer vision algorithms were developed in Python using the Spyder integrated development environment (IDE). These algorithms all complemented each other in characterizing nanomaterial thickness (*16*): imagedetector.py and thicknesscalculator.py helped calculate the layer number based on image contrast, while visualizer.py facilitated the visualization of a specific area's layers (Fig. 3).

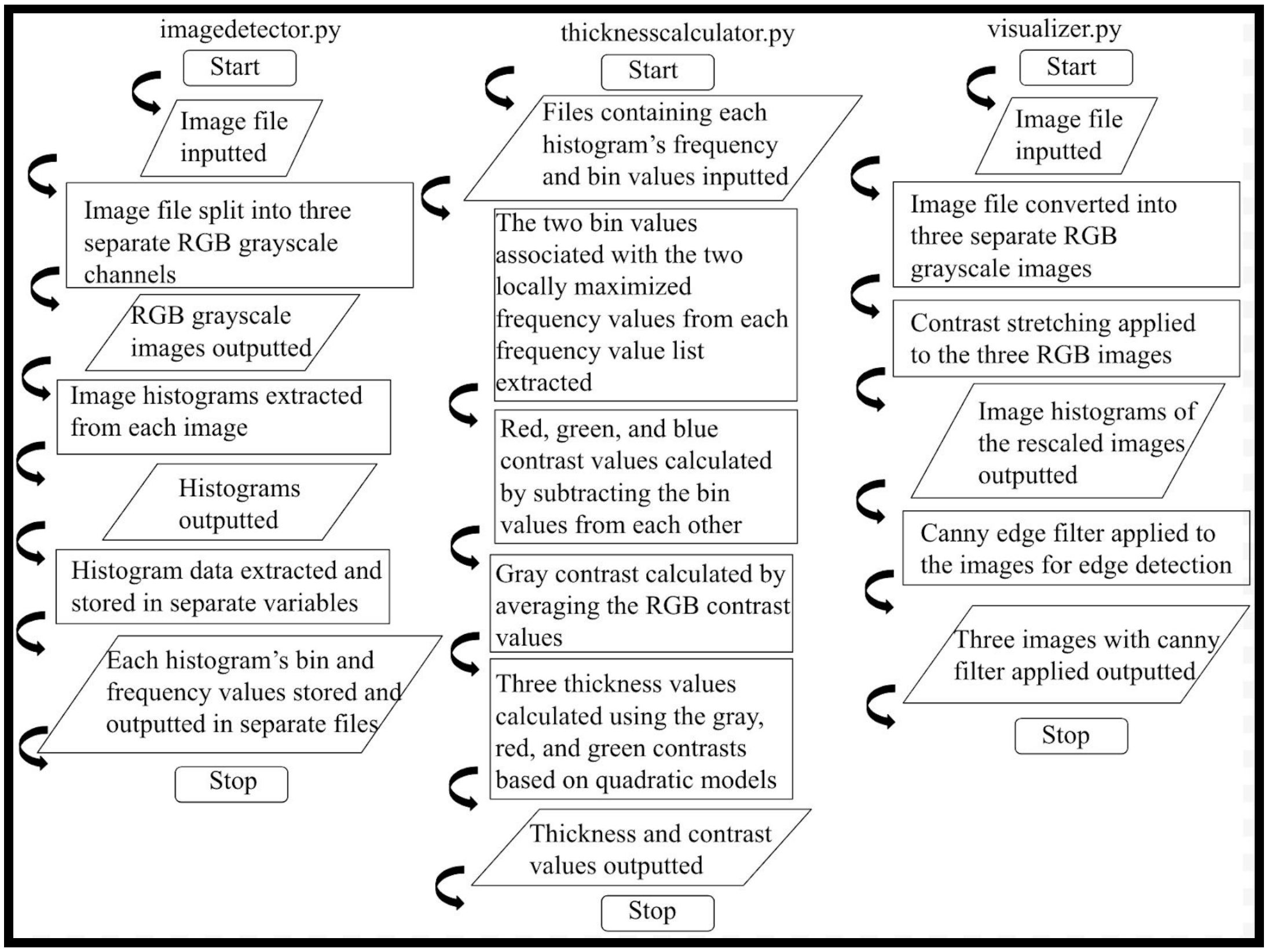

**Figure 3**: Three flowcharts summarizing each algorithm's general functionality. Both imagedetector.py and thicknesscalculator.py calculated the number of a uniform region, while visualizer.py aided in highlighting the locations and shapes of each of the layers in a sample. Though it is not shown in the figure, two different thicknesscalculator.py programs were used for $MoTe_2$ and graphene called thicknesscalculatorMoTe2.py and thicknesscalculatorgraphene.py in order to input different coefficients for the quadratic models.

In order to create the two algorithms for calculating layer number based on image contrast, quadratic models were utilized to quantitatively determine the relationship between layer number and color contrast by curve fitting the data for the 12 $MoTe_2$ samples and 16 graphene samples from another study (*9*) through the RStudio IDE platform. Once the quadratic equations that modeled layer number V.S. color contrast for the gray, red, and green grayscale color channels were acquired, the imagedetector.py and thicknesscalculator.py programs were created. The quadratic model for the blue color channel was not obtained due to its irregular and erratic contrast values.

In principle, imagedetector.py functioned by taking the digitally colored image of one of the sample's uniform regions as an input; converting it into red, green, and blue grayscale images; obtaining each image's histogram that binned pixel intensity values according to frequency of occurrence; and then separately outputting bin and frequency values from each of the three image histograms into files. These files then served as the inputs for thicknesscalculator.py in which the two bin values associated with the two locally maximized frequency values from each histogram's data set were acquired by manipulating the histogram data (Fig. 4). The difference between the two bin values, representing the pixel intensity values of the substrate and sample, were taken for the red, green, and blue grayscale channels in order to calculate the corresponding color contrasts. These contrasts were then averaged to acquire the gray color contrast. Then, the gray, red, and green contrasts were inputted into the quadratic equations obtained earlier on through RStudio to find the layer number values. In the end, the layer numbers and color contrasts were outputted in the Spyder console so that the layer numbers could be averaged to provide a more accurate measurement of the sample region's layer number. If there was an imaginary solution to one of the grayscale channel's quadratic equations, the string "4 or thicker" or "13 or thicker" would be outputted in the Spyder console if the sample was $MoTe_2$ or graphene respectively. Due to the erratic and inconsistent nature of the blue color channel, the blue color contrast was not used to calculate the layer number.

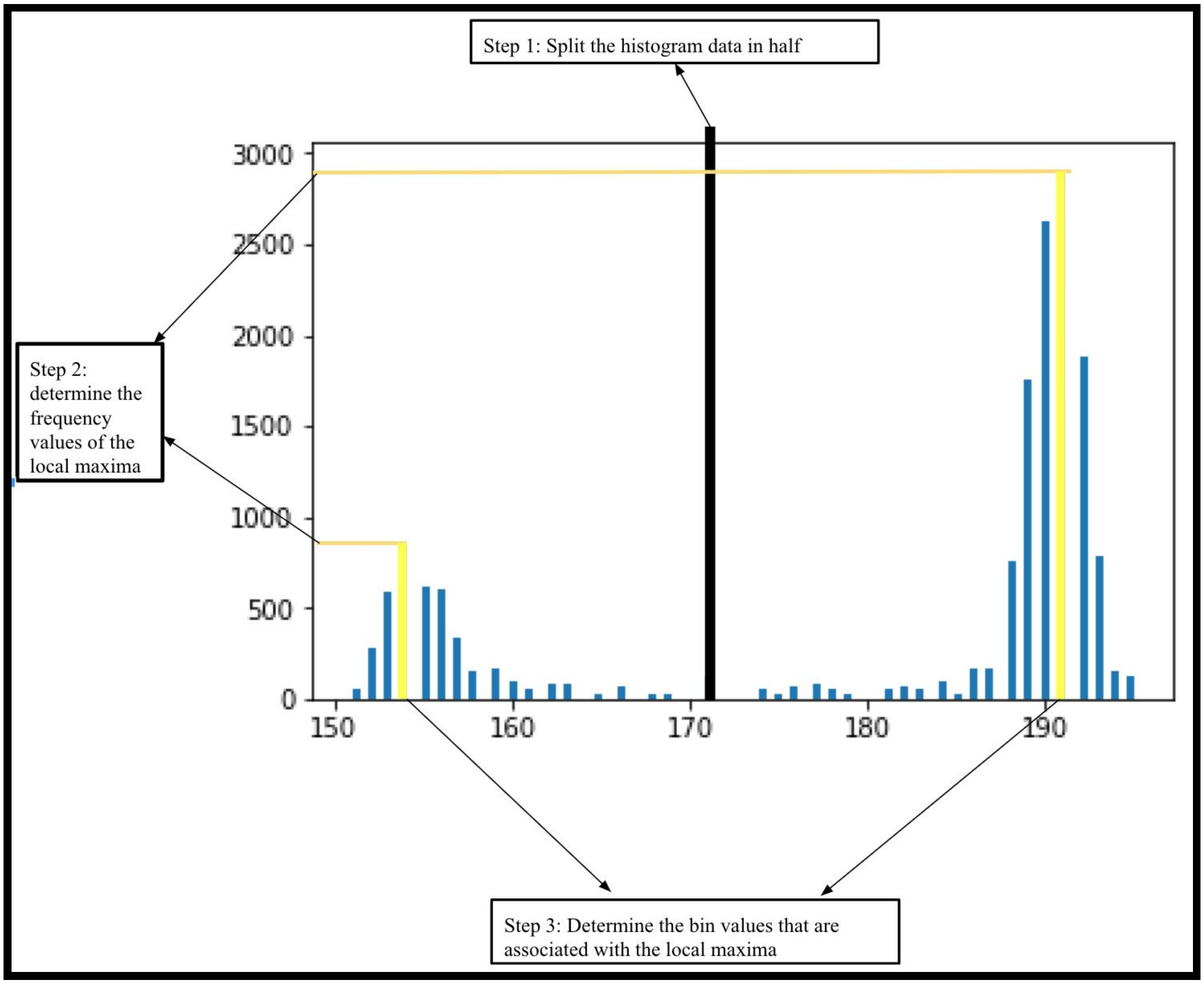

**Figure 4**: Diagram explaining the functionality of thicknesscalculator.py in which the bin values of the two locally maximized frequency values were obtained by splitting the histogram data in half, determining the maxima of the two halved lists of frequency values, and then extracting the two bin values that corresponded to the two local maxima.

Additionally, visualizer.py was mainly used as a visualization tool for highlighting the outlines of the different layers of each sample. Similar to imagedetector.py and

thicknesscalculator.py, the digital color image of the sample was inputted and split into separate gray, red, and green grayscale images. Contrast stretching (*17*), a form of histogram equalization, was applied to each of the grayscale images for all pixel intensity values that fell between the third and ninety-seventh percentiles in order to enhance the pixel intensity values, especially between two seemingly indistinguishable layers. Then a Canny filter (*17*) was applied with a filter value of between 0 and 10 in order to perform a Canny edge detection algorithm that ultimately outputted highlighted outlines of the layers for each of the three gray, red, and green grayscale images.

## Results and Discussion

### *Experimental Results and Discussion*

After executing the experimental method for all 8 of the $MoTe_2$ samples, 12 unique color contrast signatures were obtained for 12 different regions for each $MoTe_2$ crystal. While some samples were uniform with only one layer number value represented, others were composed of regions that differed in layer number, causing the optical images to look like steps. Furthermore, according to the progression of contrast values within each of the gray, red, and green grayscale channels (Fig. 5), darker and thicker regions of the samples appeared to be correlated with higher contrast values since the difference between the sample and substrate's pixel intensity values was greater than those of thinner regions (Table 2). For instance, the gray, red, and green contrast values for S4Dthinner, the pale purple square-like region on sample S4D that was measured to be 1 layer thick according to the AFM (*13*) and Gwyddion analysis (*14*), were 10.5, 8.4, and 17.9 respectively while S4Dthicker's values were 27.64, 32.8, and 47.22 for a sample that was measured to be 3 layers thick. Furthermore, not all of the color contrast values increased consistently and uniformly as the the sample's layer number increased in thickness. For example, S5Gthicker's green color contrast was 14.9 which matched the green color contrast of sample S5Gthinner despite the fact that S5Gthicker visually seemed thicker than S5Gthinner. As a consequence, sources of error could be traced to either disturbances in the vacuum chamber such as vibrations that caused the optical microscope to take blurry images of the sample or imperfections in ImageJ's software (*12*) that caused the plot profile function to miscalculate the pixel intensities. Likewise, similar sources of error could be applied to the thickness measurements of the $MoTe_2$ samples since vibrations within the AFM's proximity could have caused substantial amounts of noise in the AFM image or defects in the Gwyddion SPM analysis software could have resulted in inaccurate measurements of the layer thickness.

| | S4D * | S5D1 * | S4B1 | S5D 1 * | S4B 2 | S5G * | S4D * | S3D * | S5G | S5E | S3B * | S3D * |
|---|---|---|---|---|---|---|---|---|---|---|---|---|
| Gray | 10.5 | 11.5 | 22.6 | 20.0 | 24 | 24.5 | 27.64 | 26.25 | 28.5 | 28.5 | 29 | 28.24 |
| Red | 8.4 | 5.4 | 23.7 | 10.7 | 25.4 | 14.9 | 32.8 | 28 | 14.9 | 24.7 | 35.4 | 36.96 |
| Green | 17.9 | 16.6 | 33.2 | 33.9 | 42.7 | 39.01 | 47.22 | 45.8 | 39.01 | 48.16 | 49.8 | 54.4 |

**Figure 5**: Gray, red, and green contrast values for each of the 12 regions for 8 different $MoTe_2$ samples. Yellow represents monolayer regions, green represents bilayer samples, blue represents trilayer samples, and purple represents tetralayer samples. Sample names in the topmost row with asterisks were those that were used in the quadratic curve fitting process described in the "Computational Materials and Methods" section.

| S4Dthinner | S5D1 | S4B1 | S4B2 | S5D1thinner | S5Gthinner |
|---|---|---|---|---|---|
| 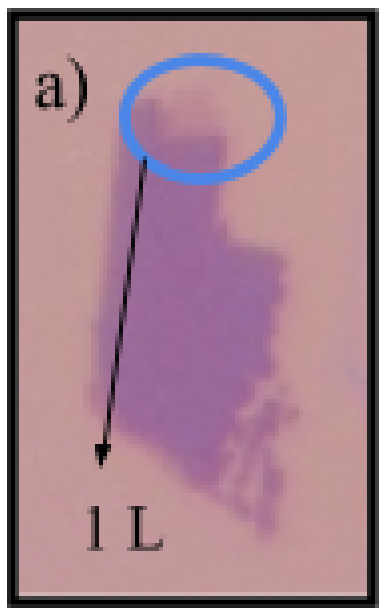  | 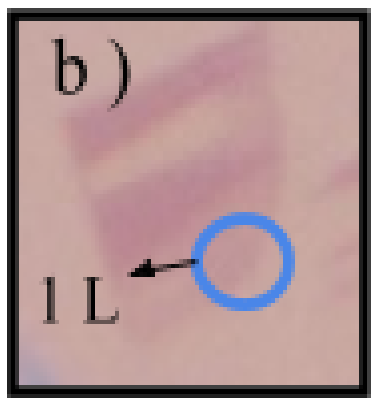  | 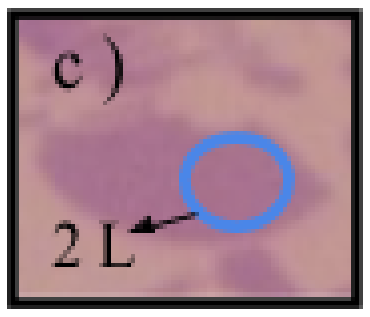  | 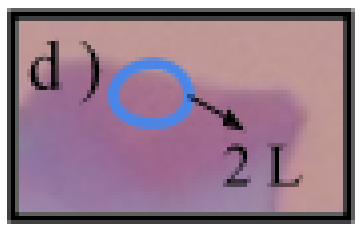  | 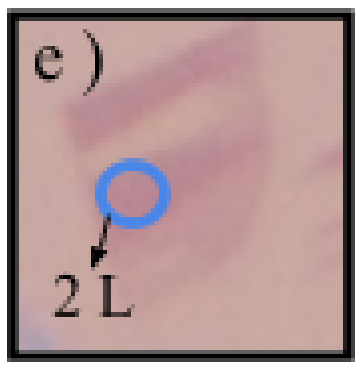  | 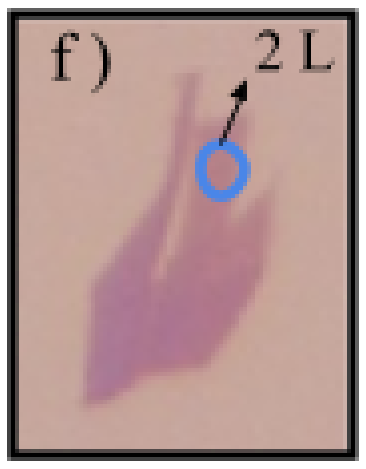  |
| S3Dthinner | S4Dthicker | S5E | S5Gthicker | S3B | S3Dthicker |
| 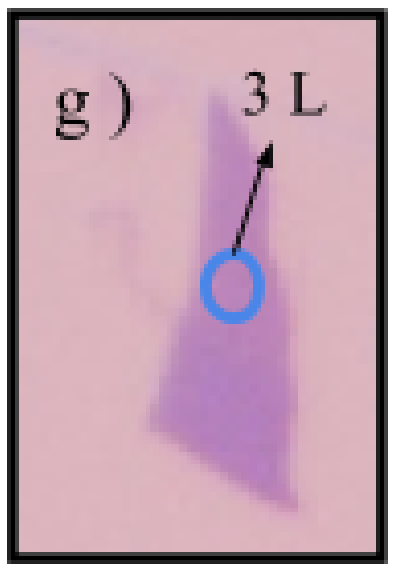  | 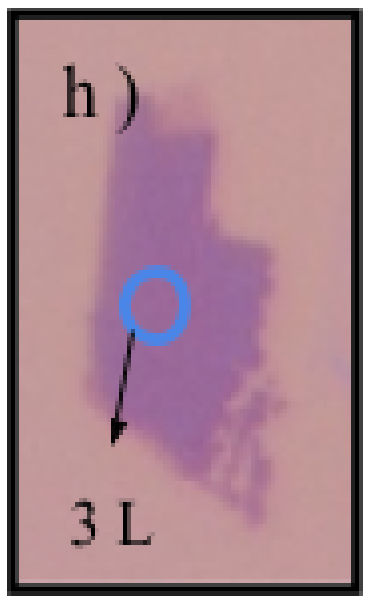  | 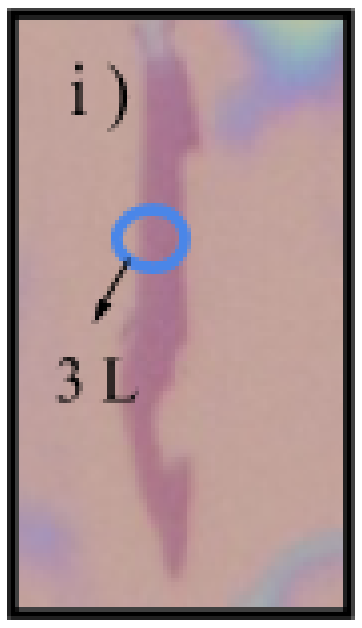  | 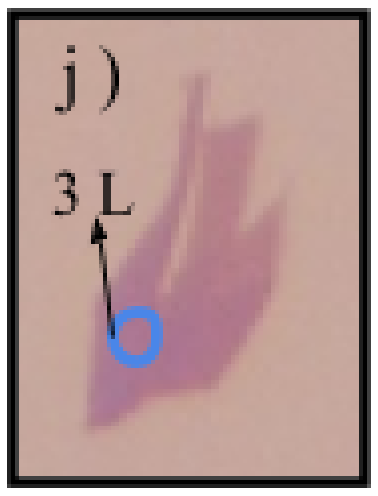  | 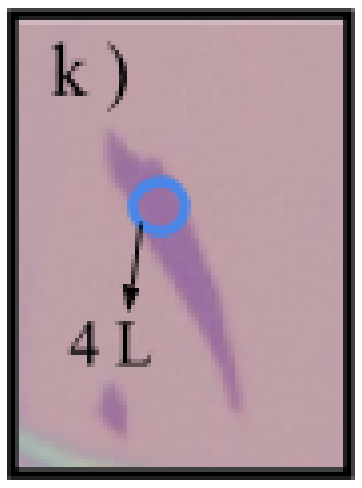  | 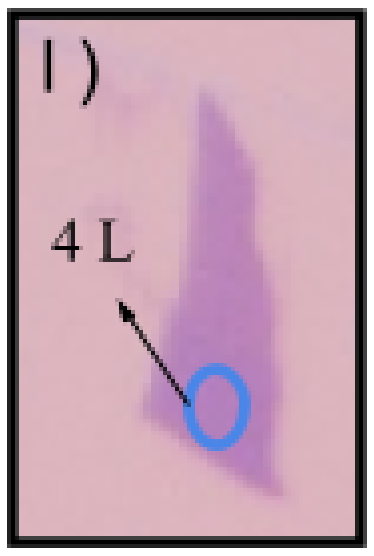  |

**Table 2**: digital color images taken on 300 nm silicon substrate and with a 50 ms exposure time of 12 regions that are circled in blue and labeled with the layer number in order from thinnest to thickest. The extensions “thinner” and “thicker” were used to indicate the relative thicknesses of regions on one sample.

*Computational Results and Discussion*

Using the $MoTe_2$ and graphene nanosheet layer number and contrast data obtained from 300 nm silicon substrate and a 50 ms exposure time, 6 quadratic models were produced for the gray, red, and green grayscale channels, three for the $MoTe_2$ samples and 3 for the graphene samples, with P-values less than 0.01000 and Pearson's multiple R-squared values greater than 0.9000 (Table 3). Moreover, the coefficients of the six quadratic models were inputted into thicknesscalculator.py so that 28 test trials of the optical-image based layer number calculator algorithm comprised of imagedetector.py and thicknesscalculator.py could be completed. As a consequence, 25 out of the 28 trials correctly calculated the layer number based on the optical image contrast after the three layer numbers were averaged, resulting in an approximate success rate of 89% (Tables 4a, 4b, and 4c). Three trials, one for a $MoTe_2$ sample and two for two graphene samples, incorrectly calculated the layer number based on the layer number that corresponded to the region using the experimental method. Therefore, potential sources of error could have been the same ones outlined in the "Experimental Results and Discussion" section that formed a faulty basis for evaluating the computer vision results. However, imperfections in the execution of the Python code and the standard error present in the 6 quadratic models could have been substantial contributors to the inaccurate measurements of the layer number based on optical image contrast. Overall, the combined execution time of imagedetector.py and thicknesscalculator.py was about 10 seconds.

<table>
<tr><th>Channels/ Sample Identity</th><th>$MoTe_2$ (5 degrees of freedom)</th><th>Graphene (10 degrees of freedom)</th></tr>
<tr><td>Gray</td><td>Equation: $- 2.399(x^2) + 17.743(x) - 4.169$<br>Multiple $r^2$: 0.9528<br>P-value: 0.000209<br>Residual standard error: 1.622<br>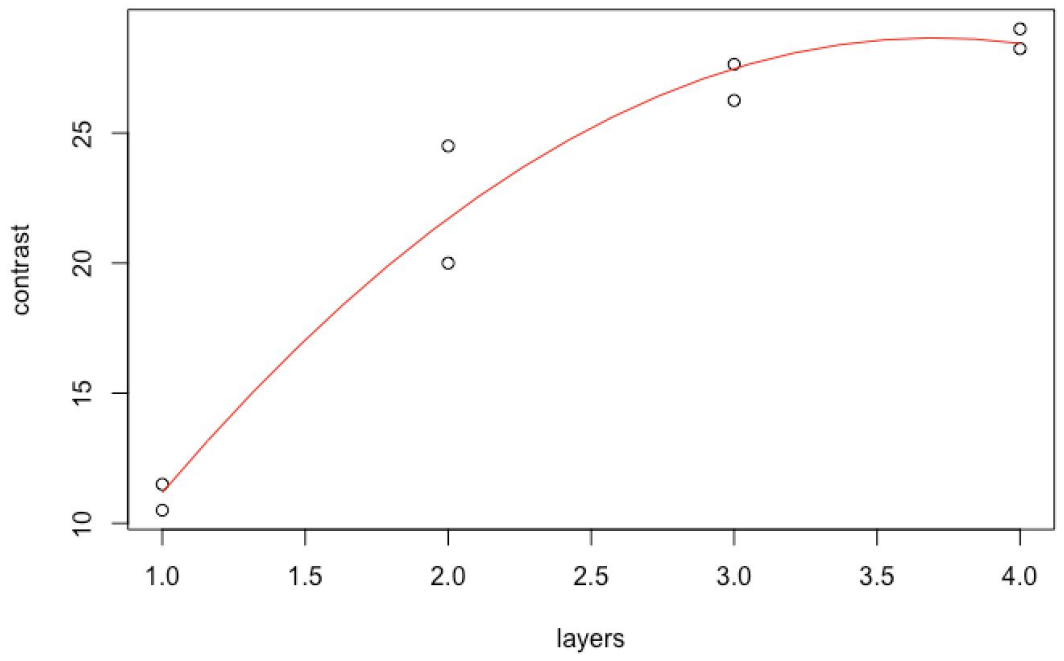
</td><td>Equation: $0.090 (x^2) - 3.817 (x) + 0.549$<br>Multiple $r^2$: 0.9979<br>P-value: $3.662 * 10^{-14}$<br>Residual standard error: 0.4973<br>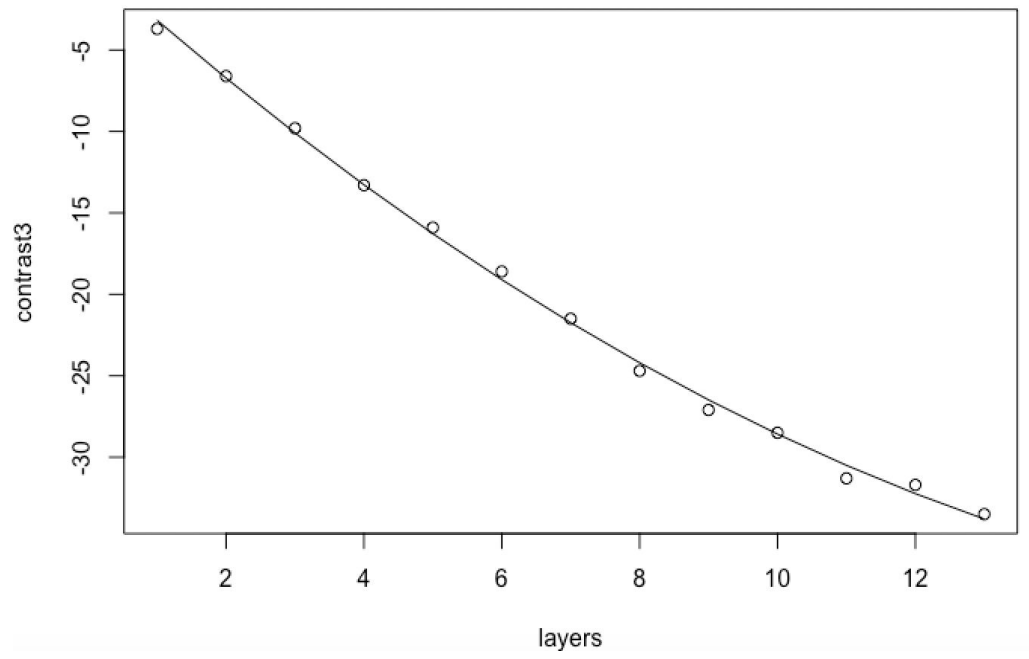
</td></tr>
<tr><td>Red</td><td>Equation: $- 0.03 (x^2) + 10.69 (x) - 4.94$<br>Multiple $r^2$: 0.9318<br>P-value: 0.001215<br>Residual standard error: 4.034<br>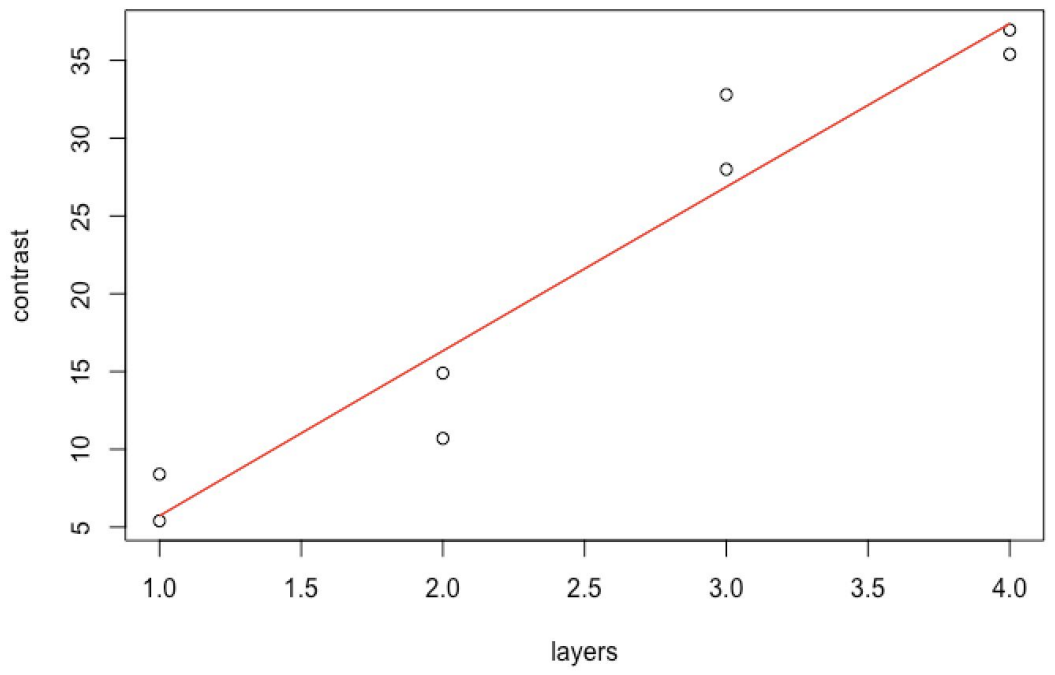
</td><td>Equation: $- 0.021 (x^2) - 3.895 (x) + 3.18$<br>Multiple $r^2$: 0.9875<br>P-value: $3.07 * 10^{-10}$<br>Residual standard error: 2.012<br>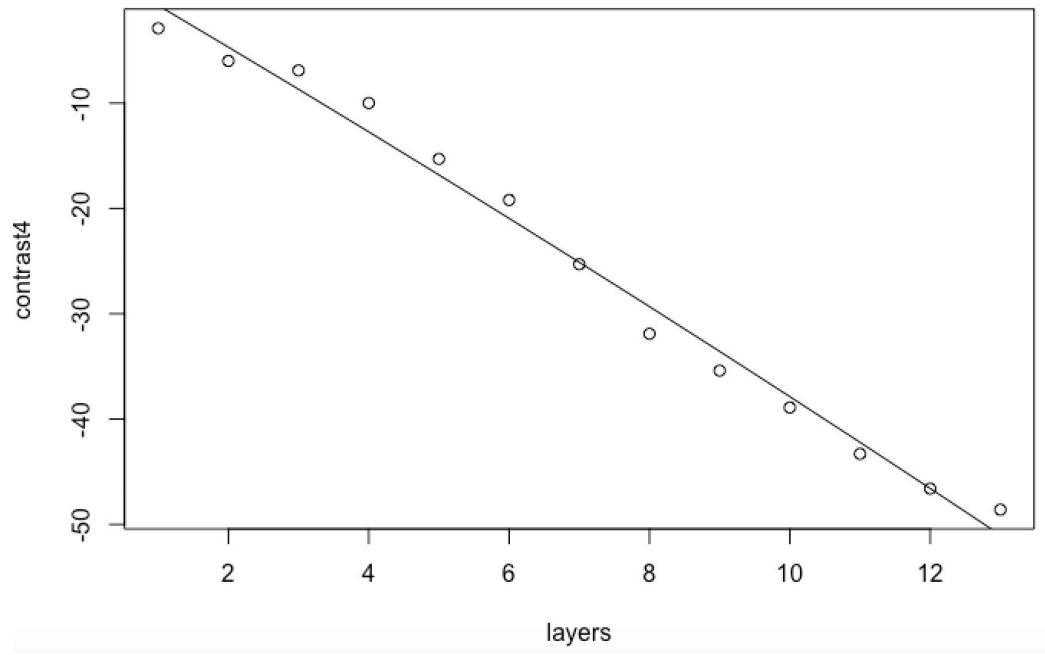
</td></tr>
<tr><td>Green</td><td>Equation: $- 3.404(x^2) + 28.479 (x) - 7.591$<br>Multiple $r^2$: 0.9807<br>P-value: $5.18 * 10^{-5}$<br>Residual standard error: 2.353</td><td>Equation: $0.324 (x^2) - 7.910 (x) - 2.775$<br>Multiple $r^2$: 0.9872<br>P-value: $1.36 * 10^{-10}$<br>Residual standard error: 1.565</td></tr>
</table>

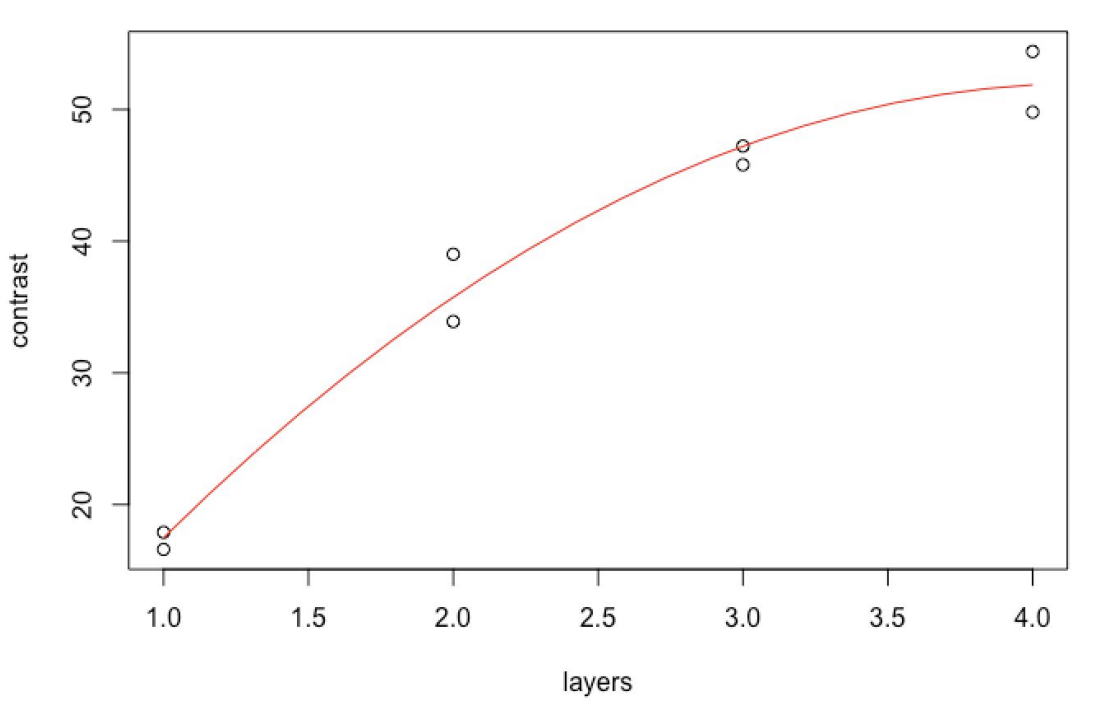

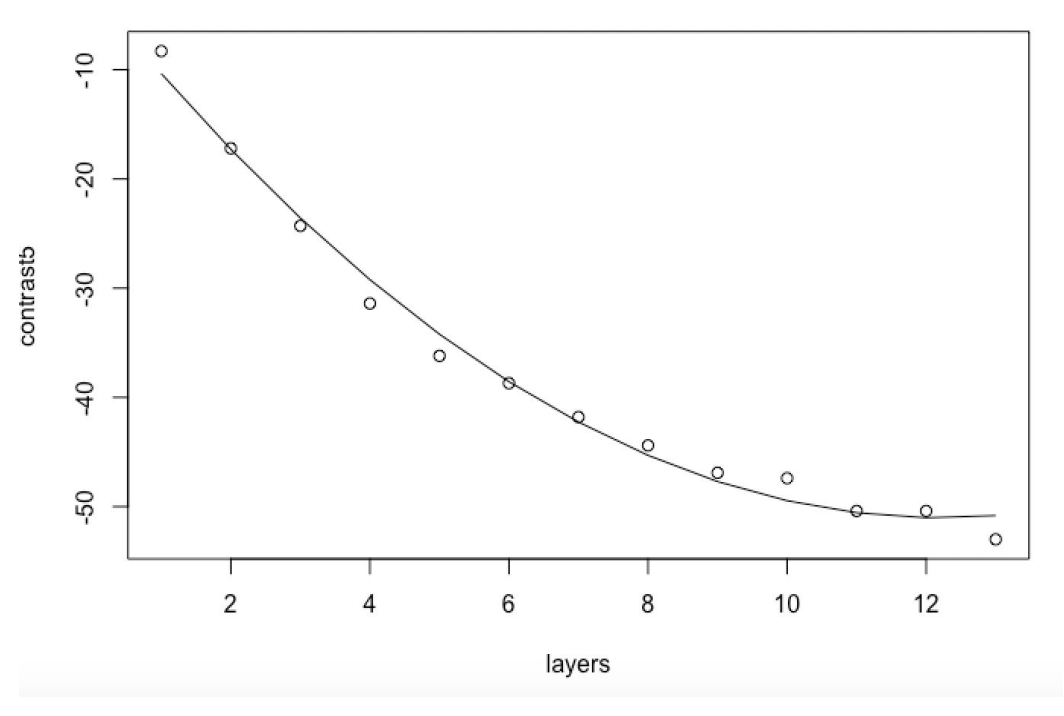

**Table 3**: 6 quadratic models of contrast V.S. layer number with their associated equation, Pearson's multiple $r^2$ value, P-value, and residual standard error for each of the gray, red, and green grayscale channels for $MoTe_2$ and graphene.

a)

| Sample Name | MoTe2 layer number | Average | Actual Value |
|---|---|---|---|
| S4Dtop | [1, 2, 1] | 1.33~1 | 1 |
| S5D1thinner | [1, 1, 1] | 1 | 1 |
| S4B1thinner | [2, 3, 2] | 2.33 ~2 | 2 |
| S4B2 | [2, 3, 2] | 2.33 ~ 2 | 2 |
| S5D1thicker | [2, 2, 2] | 2 | 2 |
| S5Gthinner | [2, 2, 2] | 2 | 2 |
| S3Dthinner | [3, 3, 3] | 3 | 3 |
| S4Dthicker | ['4 or thicker', 4, 3] | 3.5~4 | 3 |
| S5Ethicker | [3, 3, 3] | 3 | 3 |
| S5Gthicker | ['4 or thicker', 3, 3] | 3 | 3 |
| S3B | ['4 or thicker', 4, 3] | 3.5~4 | 4 |
| S3Dthicker | ['4 or thicker', 4, '4 or thicker'] | 4 | 4 |

b)

| Sample Name | Graphene layer number | Average | Actual Value |
|---|---|---|---|
| GE1 | [1, 2, 1] | 1 | 1 |
| GE2 | [3, 3, 2] | 2.67~3 | 2 |
| GE3 | [3, 3, 3] | 3 | 3 |
| GE4 | [5, 4, 4] | 4.33~4 | 4 |
| GE5 | [6, 5, 5] | 5.33~5 | 5 |
| GE6. | [7, 6, 5] | 6 | 6 |
| GE7 | [8, 7, 6] | 7 | 7 |
| GE8 | [8,7,6] | 7 | 8 |
| GE9 | [11, 9, 8] | 9.33~9 | 9 |
| GE10 | [11, 10, 8] | 10.33~10 | 10 |
| GE11 | [12, 12, 9] | 11 | 11 |
| GE12 | [12, 11, 9] | 10.67 | 12 |
| GE13 | ['15 or thicker', 13, 12] | 12.5~13 | 13 |
| GE3a | [4,3,3] | 3 | 3 |
| GE9a | [10, 9, 8] | 9 | 9 |
| GE10a | [11, 10, 8] | 9.67~10 | 10 |

c)

| | |
|---|---|
| correct | |
| incorrect | |
| total | 28 |
| correct/incorrect | 25/3 |
| Success rate | 89.30% |

**Table 4a, 4b, and 4c**: Tables showcasing averages after the layer numbers were calculated using imagedetector.py and thicknesscalculator.py. If '4 or thicker' or '13 or thicker' appeared in the output for the layer number, the value was approximated to be either 4 or 13 for $MoTe_2$ or graphene respectively since the image contrast was deemed to be close enough to the vertex of the quadratic model based on the residual standard error values. It should also be noted that $MoTe_2$ samples S4B1, S4B2, S5E, and S5Gthicker and graphene samples GE13, GE3a, GE9a, and GE10a were not used in the curve fitting process.

Lastly, visualizer.py complemented imagedetector.py and thicknesscalculator.py in order to output highlighted outlines of the layers of a sample by using a combined algorithm that was composed of contrast stretching and canny edge detection. Representing the first half of visualizer.py and a specific form of histogram equalization, contrast stretching was applied for each pixel in the image in the following equation (*16*):

$$P_{out} = (P_{in} - c)(\frac{b-a}{d-c}) + a \text{ . [1]}$$

$P_{in}$ represents the pixel from the image that is inputted, $P_{out}$ represents the pixel that is outputted after contrast stretching is applied, a and b represent the upper and lower limits of the intensity values which, for a grayscale image, are 255 and 0 respectively, and c and d represent the percentiles that are used to eliminate outliers which, for the purposes of this study, were the third and ninety seventh percentiles. Afterward, a canny edge detection algorithm (*17*), a multistage edge detector, was adapted by using a Gaussian filter that reduces noise effects by smoothing the image's edges. Furthermore the intensity gradients of the image were computed to determine horizontal, vertical, and diagonal edges in the image using the following equations

$$G = \sqrt{G_x^{\ 2} + G_y^{\ 2}}, \ \theta = arctan(\frac{G_y}{G_x}) \text{ [2], [3]}$$

Hysteresis thresholding was then applied to the intensity gradients in order to keep only the highest intensity values while thinning out the edges by removing lower intensity values.

Ultimately, the user can manipulate the sigma value that is used in computing the Gaussian filter in order to better visualize separate layers of one sample, especially those that appeared to have the same color contrast. For instance, a low sigma value allows the user to visualize the separation between two layers in sample S3D; however, once the sigma value was increased, the general outline of sample S3D became enhanced while the separation between the two layers seemed nonexistent (Fig. 6). Ultimately, visualizer.py had a total runtime of about 4 seconds.

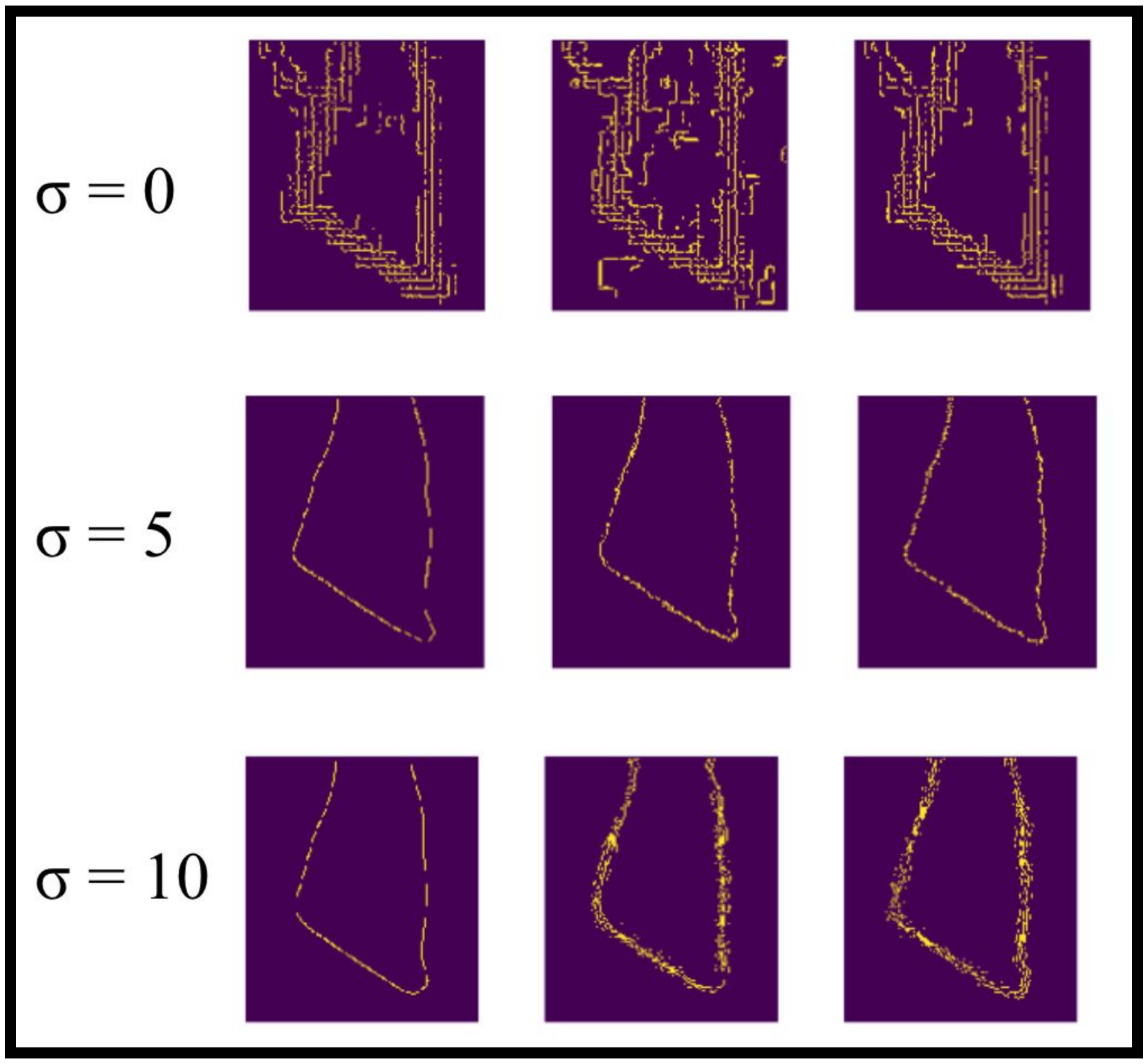

**Figure 6**: Diagram showcasing the effects of different sigma values on sample S3D's outputted images for the Canny edge detector. The yellow lines represent the computer's interpretation of the sample's outline based on the sigma value.

## Conclusion

Overall, this study demonstrated that computer vision-based algorithms can be effective at characterizing nanomaterials with a relatively high success rate of 89% and a fast runtime of around 15 seconds for all three algorithms combined. As a consequence, when compared with operating electron and force microscopes or attempts to correlate image contrast with sample layer number in previous studies (*8*-*11*), the computer vision method proved to be efficient and also cost-effective, potentially eliminating the aid of high-powered tools such as SEMs, TEMs, AFMs, and multiple beam interferometry. Therefore, relatively accurate and quick measurements can be made regarding the layer numbers of nanomaterials without the significant need to be wary about surrounding interferences, cost, and time. This may then permit labs to allocate their energy and resources on other more productive research objectives.

Nevertheless, improvements must be made for this method in order to improve its performance. Despite the fact that the method obtained a high success rate of 89%, there was still a 11% failure rate when compared to measurements executed using the AFM for the $MoTe_2$ and graphene samples (*9*). As a consequence, adjustments to the code or the image quality will have to be made in order to enable more accurate thickness characterizations. For instance, taking greater resolution pictures using the optical microscope would be able to lessen the pixelation that was observed when the images were cropped and zoomed in. Alternatively, machine learning or convoluted neural networks could be applied in conjunction with the computer vision algorithms to eliminate the need to constantly revert back to RStudio when a new data point needs to be added in order to modify the curve fitting models. Therefore, this addition to the code may be able to render the overall method more efficient and accurate in its ability to measure nanomaterial thicknesses. Furthermore, variables such as exposure time and substrate thickness were kept constant for the purposes of this study but must be varied in order to determine the overall effectiveness of the computer vision method.

However, there appears to be great promise in harnessing the power of computer vision and image processing to characterize nanomaterials. In the future, computers may be able to carefully characterize and select materials for research and industry when, for instance, a new semiconductor needs to be made. Though it may take a while before computers can completely replace humans, a realistic and substantial development of this work would be to incorporate the computer vision code with a motorized optical microscope so that layer number measurements could be made in real time while the microscope autonomously scanned over the surface of the sample. This would further enhance the characterization process's speed by integrating the search for suitable samples with thickness measurements. With these advantages, improvements, and applications in mind, more work will inevitably have to be done in order to render this method more practical for widespread use.

**Acknowledgements**

I would like to thank the Yan Optoelectronics Lab at the University of Massachusetts Amherst headed by assistant professor of physics Dr. Jun Yan for their unwavering support in mentoring me for this project and allowing me to use their resources. Specifically, I would like to thank Tom Goldstein, graduate student in the UMass physics department, and Dr. Ivory Hills, academic dean at Deerfield Academy and former research scientist at Merck Pharmaceuticals, who both gave me instrumental advice for this study. Most importantly, I also would like to thank my friends, teachers, and family for their help and support in my endeavors.